\documentstyle[epsfig,aps,prl,psfig,floats]{revtex}

\begin{document}

\twocolumn[\hsize\textwidth\columnwidth\hsize\csname
@twocolumnfalse\endcsname
\title{The Excitation Spectrum of a Bose-Einstein Condensate}
\author{J. Steinhauer, R. Ozeri, N. Katz, and N. Davidson}
\address{Department of Physics of Complex Systems,\\
Weizmann Institute of Science, Rehovot 76100, Israel}
\maketitle

\begin{abstract}
We report the first measurement of the excitation spectrum and the static
structure factor of a Bose-Einstein condensate. The excitation spectrum
displays a linear phonon regime, as well as a parabolic single-particle
regime. The linear regime provides an upper limit for the superfluid
critical velocity, by the Landau criterion. The excitation spectrum agrees
well with the Bogoliubov spectrum, in the local density approximation. This
agreement continues even for excitations close to the long-wavelength limit
of the region of applicability of the approximation. Feynman's relation
between the excitation spectrum and the static structure factor is verified,
within an overall constant.
\end{abstract}]

The excitation spectrum of superfluid $^{4}$He gives important insights into
this superfluid{\it \cite{pines90}}. The excitation spectrum places an upper
bound on the superfluid critical velocity. Furthermore, it indicates the
types of excitations which occur in the superfluid, and reflects the
superfluid's density correlations. The excitation spectrum of a
Bose-Einstein condensate should give similar insight into this system.

The excitation spectrum is the relation $\omega (k)$, giving the energy $%
\hbar \omega (k)$ of each excitation, as a function of its wave vector $k$.
\ For excitations with wavelengths $2\pi /k$\ which are comparable to the
radius of the condensate, the spectrum is characterized by discrete
oscillatory modes, which are strongly dependent on the shape of the
condensate. \ These modes have been measured previously{\it \cite{jin96}},%
{\it \cite{mewes96}}. \ For wavelengths much shorter than the radius of the
condensate in the direction of $\vec{k}$, the excitation spectrum becomes an
essentially continuous function of $k$, which characterizes the intrinsic
bulk properties of the condensate{\it \cite{pitskw}}. \ In this work,
''excitation spectrum'' refers to this bulk regime. \ We report the first
measurement of the excitation spectrum.

Previously, for condensates with various values of the chemical potential $%
\mu $, isolated points on the excitation spectrum were measured{\it \cite
{stenger99}}, {\it \cite{stamperkurn99}},{\it \cite{inouye00}},{\it \cite
{vogels}}. \ These points, and their dependence on $\mu $, were found to be
consistent with Bogoliubov theory.

For a homogeneous condensate (constant density $n$), the zero-temperature
excitation spectrum is expected to be of the Bogoliubov form{\it \cite
{bogo47}}, which is valid as long as the wavelength of the excitation is
much greater than the $s$-wave scattering length $a$. \ The Bogoliubov
excitation spectrum is given by
\begin{equation}
\omega (k)=\sqrt{c^{2}k^{2}+\left( \frac{\hbar k^{2}}{2m}\right) ^{2}}
\label{bog}
\end{equation}

The speed of sound $c$ is given by $\sqrt{gn/m}$, where the constant $g$ is
given by $4\pi \hbar ^{2}a/m$ and $m$ is the atomic mass. \ The chemical
potential $\mu =gn$ characterizes the strength of the atomic interactions. \
In the limit of small $k$, $\omega (k)\approx ck$. \ This linear part of the
excitation spectrum corresponds to phonons. \

By definition, $\omega (k)$ is the average frequency of the dynamic
structure factor{\it \cite{pines90}} $S(k,\omega )$, which gives the
response of the excitation process. \ Integrating $S(k,\omega )$ over $%
\omega $ gives the static structure factor $S(k)$, which is the Fourier
transform of the density correlation function, giving the magnitude of the
density fluctuations{\it \cite{pines66}} in the fluid, at wavelength $2\pi
/k $. $\ $In general, the $f$-sum rule{\it \cite{pines90}} yields Feynman's
relation{\it \cite{feynman}}
\begin{equation}
S(k)=(\hbar k^{2}/2m)/\omega (k)  \label{sofk}
\end{equation}
Eq. \ref{sofk} relates the strength of the resonance to its frequency.

For a BEC in a parabolic trap, the density is inhomogeneous, so (\ref{bog})
does not precisely apply. \ However, we will see that the form of $\omega
(k) $ is very close to that of (\ref{bog}). \ In the local density
approximation{\it \cite{stamperkurn99}} (LDA), the speed of sound, and
therefore the excitation spectrum (\ref{bog}), can be considered to be
defined at each point $\vec{r}$ in the BEC, where $c=\sqrt{gn(\vec{r})/m}$.
\ This approximation is valid as long as the Thomas-Fermi radius of the
condensate in the $\vec{k}$-direction is much larger than the wavelength of
the excitation {\it \cite{pitskw},\cite{stamperkurn99}},{\it \cite{comment5}%
, \cite{stringari98}}. \ For the parabolic density profile of the condensate
in the Thomas-Fermi regime, the excitation spectrum for the entire
condensate in the LDA is
\begin{equation}
\omega (k)=\sqrt{c_{ld}^{2}(k)k^{2}+\left( \frac{\hbar k^{2}}{2m}\right) ^{2}%
}  \label{lda}
\end{equation}
where $c_{ld}(k)$\ is given by $\frac{\hbar k}{2m}\sqrt{S(k)^{-2}-1}$, and
{\it \cite{stamperkurn99}},{\it \cite{pitskw}}
\begin{equation}
S(k)=\frac{15}{4}\left\{ \frac{3+\alpha }{4\alpha ^{2}}-\frac{3+2\alpha
-\alpha ^{2}}{16\alpha ^{5/2}}\left[ \pi +2\arctan \left( \frac{\alpha -1}{2%
\sqrt{\alpha }}\right) \right] \right\}  \label{sofklda}
\end{equation}
where $\alpha \equiv 2\mu /[\hbar ^{2}k^{2}/(2m)]$ and $\mu =gn(0)$, where $%
n(0)$ is the maximum density in the condensate. \ The value $c_{ld}\left(
k\right) $\ is a weak, monotonically increasing function, which varies from $%
c_{eff}\equiv 32/(15\pi )\sqrt{\mu /m}(=0.68\sqrt{\mu /m})$ for small $k$,
to $c_{\text{{\it large}}}=\sqrt{4/7}\sqrt{\mu /m}(=0.76\sqrt{\mu /m})$ for
large $k$. \ The value $c_{\text{{\it large}}}$\ is just the speed of sound
for a homogeneous condensate whose density is the density of the
inhomogeneous condensate, averaged over each atom. \ Since $c_{ld}(k)$ is
nearly constant, the excitation spectrum in the LDA (\ref{lda}) is very
similar in form to the homogeneous excitation spectrum (\ref{bog}),
illustrating that the excitation spectrum of the trapped condensate should
largely reflect the intrinsic properties of the homogeneous condensate. \ \
Thus, the functional form of $\omega (k)$ reported here provides a test for
the Bogoliubov theory for a homogeneous condensate.

Since small $k$ corresponds to phonons, $c_{eff}$\ is the effective speed of
sound for an inhomogeneous condensate. \ The Bogoliubov spectrum in the LDA
fulfills the requirement of Bose statistics that the lowest lying
excitations be phonons {\it \cite{huang}}. \ Phonons are collective
excitations, each of which consists of a large number{\it \cite{pitbog}} $%
N_{k}$ of atoms moving with momentum $\hbar k$, and $N_{k}-1$ atoms moving
with momentum -$\hbar k$. \ This momentum distribution of a phonon was
measured in {\it \cite{vogels}}. \ Previously, a sound pulse, composed of a
combination of phonons, was imaged in a time sequence of absorption images%
{\it \cite{andrews97}}. \ The observed sound velocity was in rough agreement
with $c_{eff}$.

For large $k$, (\ref{lda}) is given by
\begin{equation}
\omega (k)\approx \hbar k^{2}/(2m)+mc_{\text{{\it large}}}^{2}/\hbar
\label{largek}
\end{equation}
where the first term is much larger than the second. This parabolic part of
the spectrum corresponds to single-particle excitations, in which a single
atom has a velocity $\hbar k/m$, which is much larger than the speed of
sound. \ The second term in (\ref{largek}) is equal to $(4/7)\mu /\hbar $
and is independent of $k$, reflecting the extra interaction energy{\it \cite
{stenger99}} experienced by a moving atom{\it \cite{leggett}}.

The transition from collective excitations to single-particle excitations
occurs for $k$ on the order of $\xi ^{-1}$, where $\xi $ is the healing
length{\it \cite{baym96}}. \ We take $\xi ^{-1}$\ to be the solution of $k=%
\sqrt{2}mc_{ld}(k)/\hbar $, which is given by $\xi ^{-1}=\sqrt{2}m\overline{c%
}/\hbar $, and $\overline{c}\approx (c_{eff}+c_{\text{{\it large}}})/2$.

\ An upper limit on the superfluid critical velocity can be extracted from
the excitation spectrum, via the Landau criterion, which states that the
superfluid cannot flow with a speed greater than $\omega /k$, for any
excitation $\omega (k)$ in the spectrum{\it \cite{pines90}}. \ Therefore,
the critical velocity $v_{c}$ cannot be more than the smallest value of $%
\omega /k$ in the entire spectrum. \ According to Bogoliubov theory, the
phonons have the smallest value of $\omega /k$, as seen in Fig. 2a, so $%
v_{c} $ cannot be greater than the speed of sound $c_{eff}$. \ We emphasize
that the critical velocity for vortex production is usually much lower than
the critical velocity given by the Landau criterion, so vortex production
usually limits the speed of superfluid flow{\it \cite{raman99}},{\it \cite
{4HeVortex}}.

The condensate of $^{87}$Rb atoms in the 5s$_{1/2}$, $F=2$, $m_{F}=2$ ground
state is produced in a QUIC magnetic trap{\it \cite{esslinger98}}, loaded by
a double MOT\ system. \ The magnetic trap contains 6$\times $10$^{7}$ atoms.
\ After 22 seconds of evaporation, 1$\times 10^{5}$ atoms remain, forming a
nearly pure condensate, with a thermal fraction of 5\% or less. \ The bias
magnetic field is 2 G. \ The radial and axial trapping frequencies are 25 Hz
and 220 Hz respectively, \ yielding radial and axial Thomas-Fermi radii of 3
$\mu m$ and $R=28$ $\mu m$, respectively. \ The chemical potential $\mu
/h=2.02\pm 0.09$ kHz is determined{\it \cite{castin96}} by the radial size
of the condensate in a time-of-flight image.

The excitation spectrum is measured by Bragg spectroscopy{\it \cite
{stenger99}},{\it \cite{kozuma99}}. \ Two Bragg beams $A$ and $B$ with
approximately parallel polarization{\it \cite{inouye99}}, separated by an
angle $3^{\text{o}}\leq \theta $ $\leq 130^{\text{o}}$, illuminate the
condensate for a time $t_{B}$. \ The frequency of beam $A$ is greater than
the frequency of beam $B$ by an amount $\omega $ determined by two
acousto-optic modulators. \ If a photon is absorbed from $A$ and emitted
into $B$, an excitation is produced with energy $\omega $ and momentum $k$,
where $k=2k_{p}\sin \left( \theta /2\right) $, and $k_{p}$ is the photon
wave number. \ Here, we neglect the possibility that a single photon will
excite multiple excitations, in contrast to the case of superfluid $^{4}$He,
in which multiparticle excitations are an important contribution{\it \cite
{griffin93}},{\it \cite{pines90}} to $S(k,\omega )$.

The wave vector $\overrightarrow{k}$ is precisely adjusted to be along the
axis of the cigar-shaped condensate{\it \cite{comment4}}. \ To insure that
the entire condensate is stimulated by the Bragg pulse, the length of the
pulse $t_{B}$ is chosen such that the spectral width of the pulse is roughly
equal to the intrinsic width of the resonance. \ For this experiment, the
broadening due to inhomogeneous density $\Delta \nu _{ld}$ always dominates
the Doppler broadening{\it \cite{stenger99}}, and is given by{\it \cite
{pitskw}} 0.45 kHz for large $k$, and 0.3 $\omega (k)/(2\pi )$ in the phonon
regime. \ Thus, $t_{B}$ is chosen to be roughly $(2\Delta \nu _{ld})^{-1}$.
\ For large $k$, the resonance may be further broadened by $s$-wave
scattering{\it \cite{comment6}}.

The beams are detuned $\Delta =6.5$ GHz below the 5S$_{1/2}$, $F=2$ $%
\rightarrow $ 5P$_{3/2}$, $F=3$ transition. \ The intensities $I_{A}$ and $%
I_{B}$ of each beam are adjusted to values between 0.1 mW cm$^{-2}$ and 1.1
mW cm$^{-2}${\it \cite{comment3}}, so that the number of excitations is 10\%
to 20\% of the number of atoms in the condensate. \ For pulses of this
strength, the chemical potential decreases by an average of only 12\% during
the pulse, so the Bragg scattering can be considered to be a small
perturbation to the condensate. The time average of $\mu /h$ during the
pulse is $1.91\pm 0.09$ kHz, which is taken to be the relevant value for the
excitation spectrum.

To measure a single point on the excitation spectrum $\omega (k)$, $k$ is
fixed by $\theta $, and $\omega $ is varied. \ The resonant value of $\omega
$ is taken as $\omega (k)$.

After the Bragg pulse, the atoms are allowed to expand freely, after which
they are imaged by absorption{\it \cite{comment2}}, as shown in Fig. 1 for $%
k $ within the phonon regime. \ Two adjacent clouds are seen. \ The left
cloud corresponds to the condensate. \ The right cloud, displaced in
velocity space by $\hbar k/m$, corresponds to the excitations. \ During the
time-of-flight, the density of the atomic cloud decreases dramatically,
resulting in a decrease in the speed of sound. \ The phonons of wave number $%
k$ are thus transformed{\it \cite{stamperkurn99}} into free particles, which
are subsequently imaged. \ The release process slightly distorts the
excitation cloud by mean-field repulsion. \ This distortion may also contain
information regarding the excitation energy.

\begin{figure}[h]
\begin{center}
\mbox{\psfig{figure=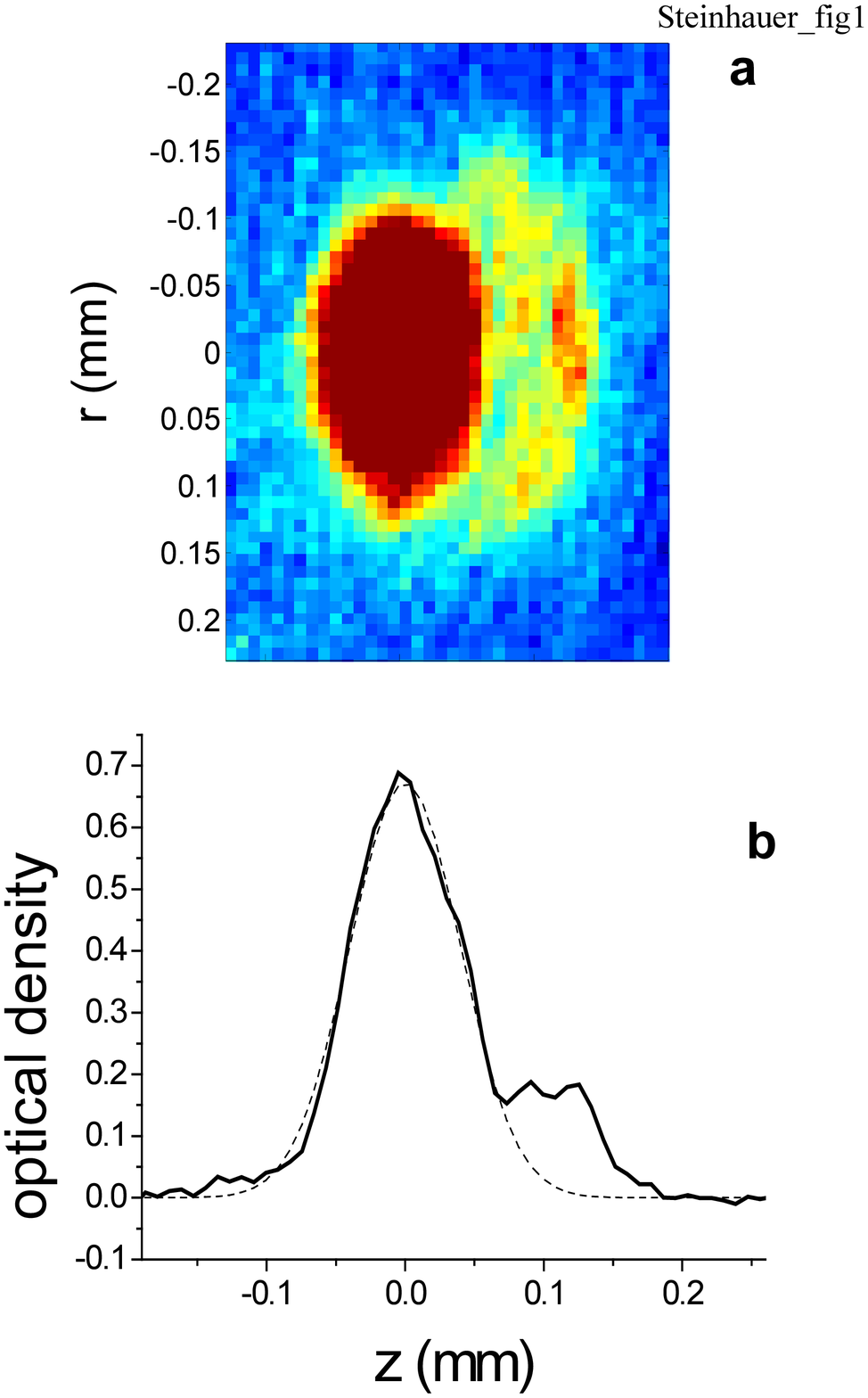,width=5cm}}
\end{center}
\vspace{0.4cm} \caption{The Bragg and condensate clouds. \ (a)
Average of 2 absorption images after 38 msec time-of-flight,
following a resonant Bragg pulse with $k=2.8$ $\mu m^{-1}$, in the
phonon regime. \ (b) \ Cross-section of the same image. \ The
dashed line is a Gaussian fit to the condensate cloud, used to
find the zero of momentum. \ The radial and axial coordinates are
indicated by $r$ and $z$, respectively.}
\end{figure}
To determine the efficiency of stimulation of excitations by the
Bragg pulse, the total momentum in the axial direction relative to
the center of the condensate cloud is computed from the image, in
the combined regions of
the two clouds{\it \cite{comment1}}. \ The total momentum is divided by $N_{%
\text{o}}\hbar k$, where $N_{\text{o}}$ is the average number of atoms in
the condensate during the Bragg pulse, to obtain the efficiency. \ This
efficiency is somewhat exaggerated though, because the total momentum
includes momentum from the release process.

The thus-measured efficiency $P(k,\omega )$ for each direction is shown in
Fig. 2, for $k=2.8$ $\mu m^{-1}$. \ The curve of $P(k,\omega )$ is
well-approximated by a Gaussian plus a constant. \ The constant results from
the background in the images. \ The symmetric shape of $P(k,\omega )$ about
the resonance frequency probably reflects the spectral shape of the Bragg
pulse, rather than the intrinsic shape which is expected to be asymmetric%
{\it \cite{pitskw}}. \ Therefore, the notation $P(k,\omega )$ is employed,
rather than $S(k,\omega )$, whose shape is the intrinsic shape.

The resonant frequency is taken as the center value of the Gaussian fit to $%
P(k,\omega )$, as shown in Fig. 2. \ The excitation energy $\omega (k)$ is
taken as the average of the resonant frequencies for the left and right
directions, which removes the effects of the Doppler shift resulting from
any sloshing of the condensate in the trap during the Bragg pulse.

\begin{figure}[h]
\begin{center}
\mbox{\psfig{figure=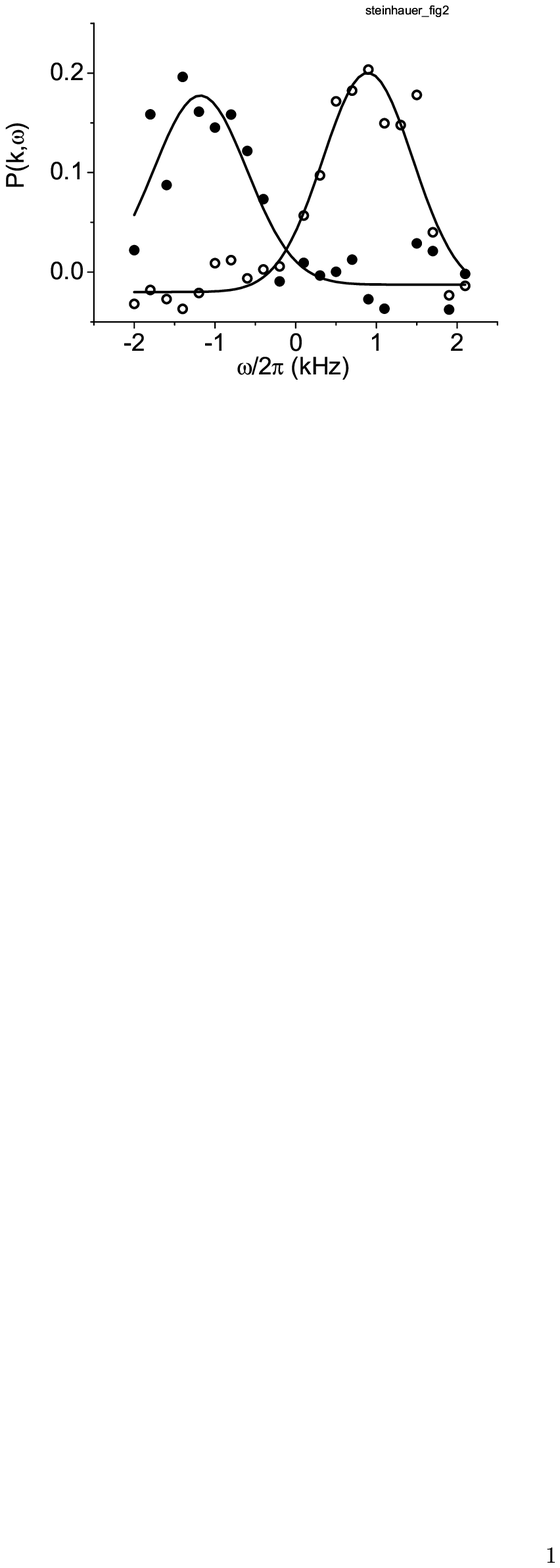,width=8cm}}
\end{center}
\vspace{0.4cm} \caption{The efficiency $P(k,\omega )$ for $k=2.8$
$\mu m^{-1}$. \ The open and filled circles are for left and
right-travelling clouds, respectively. \ The lines are fits of a
Gaussian plus a constant.}
\end{figure}
The integral of $P(k,\omega )$ over $\omega $, equal to the integral of $%
S(k,\omega )$, is related to $S(k)$ by\cite{ketterle01}
\begin{equation}
S(k)=2(\pi \Omega _{R}^{2}N_{\text{o}}t_{B})^{-1}\int P(k,\omega )d\omega
\label{rate}
\end{equation}
where $\Omega _{R}=\Gamma ^{2}/(4\Delta )\sqrt{I_{A}I_{B}}/I_{sat}$ is the
two-photon Rabi frequency, $\Gamma $ is the line width of the 5P$_{3/2}$, $%
F=3$ excited state, and $I_{sat}$ is the saturation intensity.

\ Fig. 3a shows the measured excitation spectrum. \ The solid line is the
Bogoliubov spectrum in the LDA (\ref{lda}), without any fit parameters, for $%
\mu =1.91$ kHz. \ The measured spectrum is seen to agree well with the
Bogoliubov spectrum. \ A linear phonon regime is seen for low $k$, and a
parabolic single-particle regime for high $k$. \ The line at $k=\xi ^{-1}$\
separates these two regimes. \ The excitations seen to have the smallest
value of $\omega /k$ are the phonons. \ Therefore, by the Landau criterion,
the superfluid velocity $v_{c}$ is bounded by $\omega /k$ for the phonons.

The inset of Fig. 3a shows the low $k$ region of the excitation spectrum. \
To extract the initial slope from the data, (\ref{lda}) is fit to the points
with $k$ less than 3 $\mu m^{-1}$, with $\mu $ taken as a fit parameter. \
The fit is not shown in the figure. \ The result gives the speed of sound
for the condensate to be $c_{eff}=2.0\pm 0.1$ mm sec$^{-1}$, which is also
the measured upper bound for $v_{c}$, by the Landau criterion. \ This value
is in good agreement with the theoretical value of $2.01\pm 0.05$ mm sec$%
^{-1}$, which is the $k=0$ slope of the Bogoliubov spectrum for $\mu
=1.91\pm 0.09$ kHz. \ \ The line at $2\pi R^{-1}$ indicates the excitation
whose wavelength is equal to the Thomas-Fermi radius of the condensate in
the axial direction. \ The measured excitation spectrum agrees with the LDA,
even for $k$ values approaching this lower limit of the region of validity.
\ For the two lowest $k$-values measured, the wavelength of the excitations
are long enough that they are clearly visible, as density modulations in the
condensate cloud.

\begin{figure}[h]
\begin{center}
\mbox{\psfig{figure=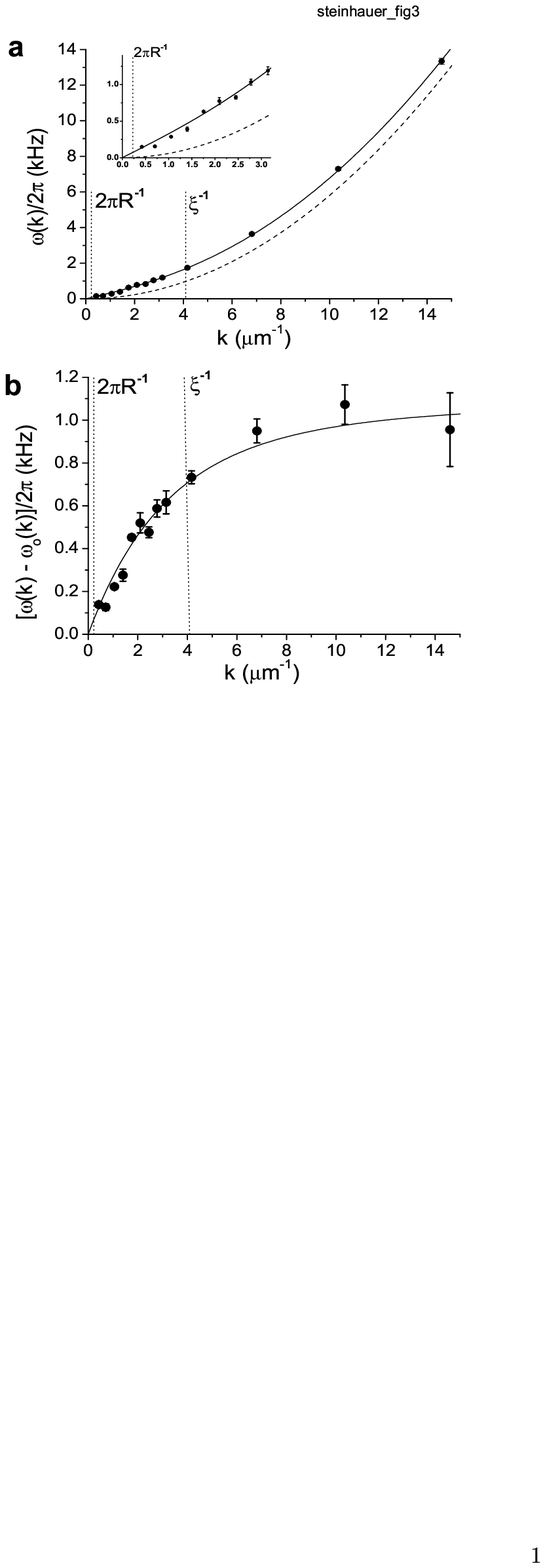,width=8cm}}
\end{center}
\vspace{0.4cm} \caption{(a) \ The measured excitation spectrum
$\omega (k)$ of a trapped Bose-Einstein condensate. \ The solid
line is the
Bogoliubov spectrum with no free parameters, in the LDA for $\mu =1.91$%
 kHz. \ The dashed line is the parabolic free-particle
spectrum. \ The vertical line at $\xi ^{-1}$ shows the separation
between the collective and single-particle regimes. \ The vertical
line at $2\pi R^{-1}$ shows the lower limit of the region of
validity of the LDA. \ For most points, the error bars are not
visible on the scale of the figure. \ The inset shows the linear
phonon regime. \ (b) \ The difference between the
excitation spectrum and the free-particle spectrum. \ Error bars represent $%
1\sigma $ statistical uncertainty. \ The theoretical curve is the
Bogoliubov spectrum in the LDA for $\mu =1.91$\ kHz, minus the
free-particle spectrum.}
\end{figure}
Fig. 3a also shows the parabolic spectrum for free particles, $\hbar
k^{2}/(2m)$. \ The measured excitation spectrum is clearly above this curve,
reflecting the interaction energy of the condensate. \ To emphasize the
interaction energy, $\omega (k)$ is shown again in Fig. 3b, after
subtraction of the free-particle spectrum. \ The theoretical curve is the
Bogoliubov spectrum in the LDA, given by (\ref{lda}), minus the parabolic
spectrum for free atoms. \ This curve approaches a constant for large $k$,
given by the second term in (\ref{largek}).

The closed circles in Fig. 4 are the measured static structure factor $S(k)$%
, by (\ref{rate}). \ The values shown have been increased by a factor of
2.3, giving rough agreement with $S(k)$ from Bogoliubov theory in the LDA (%
\ref{sofklda}). \ Eq. (\ref{sofklda}) is indicated by a solid line. \ The
required factor of 2.3 probably reflects inaccuracies in the various values
needed to compute $\Omega _{R}$. \ The open circles are computed from (\ref
{sofk}), using the measured values of $\omega (k)$ shown in Fig. 3a. \ The
rough agreement between the closed and open circles is consistent with the
relation (\ref{sofk}), within the multiplicative constant applied to the
closed circles. \ For the determination of $S(k)$, it is critical that the
apparent number of excitations is not enhanced by extra momentum obtained
during the release process. \ Therefore, the number of excitations is
determined by the number of atoms in the excitation cloud, rather than by
the total momentum. \ This technique fails for the two points with the
lowest $k$ values, where many of the atoms do not exit the condensate cloud.
\ For these points, the measured $S(k)$ seen in Fig. 4 is significantly
reduced.

\begin{figure}[h]
\begin{center}
\mbox{\psfig{figure=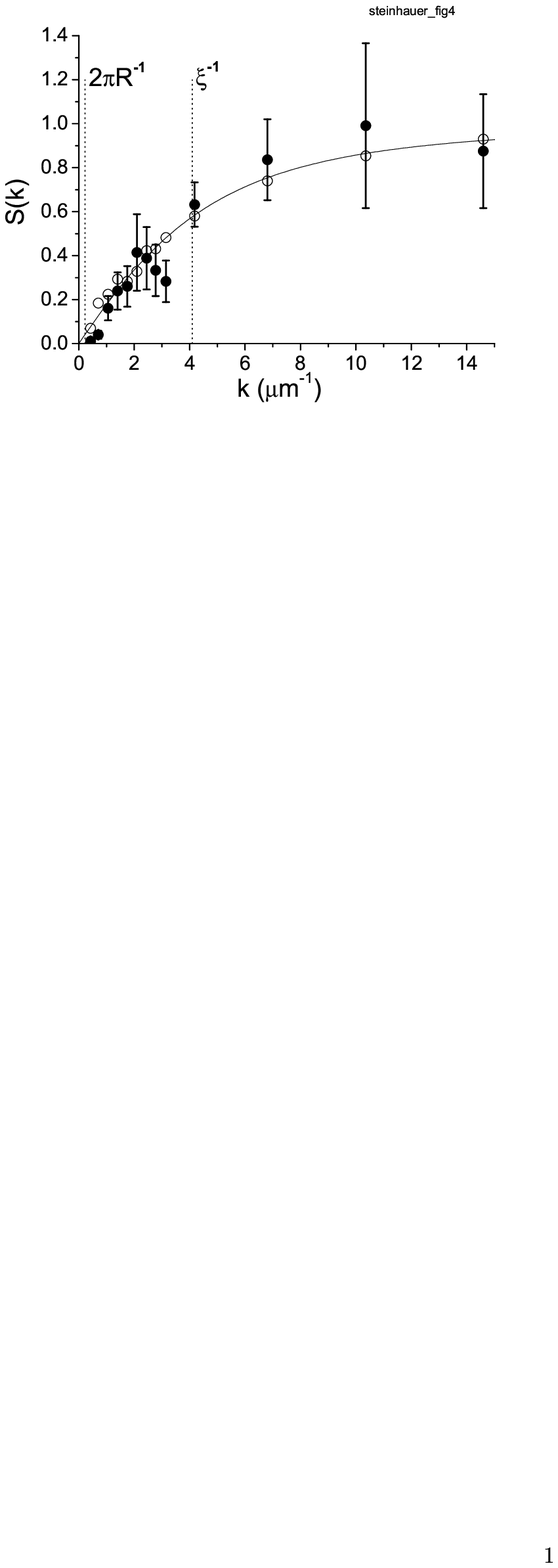,width=8cm}}
\end{center}
\vspace{0.4cm} \caption{The filled circles are the measured static
structure factor, multiplied by an overall constant of 2.3. \ \
Error bars represent $1\sigma $ statistical uncertainty, as well
as the estimated uncertainty in the two-photon Rabi frequency. \
The solid line is the Bogoliubov structure factor, in the LDA for
$\mu =1.91$ kHz. \ The open circles are computed from the measured
excitation spectrum of Fig. 3, and Feynman's relation
(\ref{sofk}). \ For the open circles, the error bars are not
visible on the scale of the figure.}
\end{figure}
For large $k$ (short wavelength), $S(k)$ approaches unity, corresponding to
non-interacting, uncorrelated atoms. \ For long wavelengths however, $S(k)$
counterintuitively approaches zero. \ For decreasing $k$, the condensate
contains increasing numbers of atoms with momentum $\hbar k$. \ These atoms,
rather than creating additional density fluctuations with wavelength $2\pi
/k $, actually suppress such fluctuations, because the atoms are correlated
in pairs with momenta $\pm \hbar k$, and opposite phase{\it \cite
{stamperkurn99}}.

Since $S(k)$ is always less than unity for the values of $k$ measured here,
the density fluctuations are never greater than in the uncorrelated case. \
However, $S(k)$ is expected to have a peak (a roton) at a wavelength
comparable to $a$, which is much shorter than the wavelengths reported here,
but the increase of $S(k)$ at this roton is on the order of $a^{3}n$. \ For
a BEC in an alkali gas, $a^{3}n\sim 10^{-4}$, so the roton is negligible. \
This is in contrast to superfluid $^{4}$He, \ where $S(k)$ has a significant
roton, corresponding to a minimum in $\omega (k)$.

For $k\geq 6.8$ $\mu m^{-1}$, $s$-wave scattering is clearly visible. \ The
measured effective scattering cross section decreases with decreasing $k$,
as predicted in {\it \cite{chikkatur}}.

In conclusion, we report the first measurement of the excitation spectrum of
a Bose-Einstein condensate, and the static structure factor. \ This spectrum
displays several features consistent with theory. \ The excitation spectrum
consists of a linear phonon regime, as well as a single-particle regime. \
The linear regime provides an upper limit for the superfluid critical
velocity, by the Landau criterion. \ The excitation spectrum agrees
quantitatively with the Bogoliubov spectrum, in the local density
approximation. \ This agreement continues even for long-wavelength
excitations, close to the limit of the region of applicability of the
approximation. \ The density fluctuations implied by the static structure
factor agree with the excitation spectrum within a multiplicative constant,
via Feynman's relation. \ Feynman's relation is thus verified within an
overall constant.


\begin{references}
\bibitem{inouye99}  Although the polarization is not along the axis of the
condensate, superradiance has not been observed in this system. \ See S.
Inouye et al., Science {\bf 285}, 571 (1999).
\bibitem{comment4}  For two points on the excitation spectrum, $k$ is not
along the axis of the condensate. $\ $Specifically, for $k=14.6$ $\mu m^{-1}$
and $k=4.2$ $\mu m^{-1}$, $\overrightarrow{k}$ is 25$^{\text{o}}$ and 4$^{%
\text{o}}$ from the axis, respectively. \ For both of these $k$-values
however, the wavelength is much smaller than the Thomas-Fermi radius in the
direction of propagation.
\bibitem{comment6}  $s$-wave scattering may cause a very small shift in $%
\omega (k)$, which we neglect.
\bibitem{comment3}  Spontaneous scattering is negligible for these
intensities.
\bibitem{comment2}  Typically, 4 images are averaged for each value of $%
\omega $ and $k$.
\bibitem{comment1}  Since the excitation cloud contains few atoms, and
overlaps the condensate, it is difficult to distinguish between atoms in the
excitation cloud and atoms in the thermal cloud. \ Therefore, the cutoff
frequency of the evaporation is adjusted for each image with a precision of
2 kHz, so that the measured thermal fraction is no more than 5\%.
\end{references}
\end{document}